\documentclass[reprint,pra, twocolumn, amsmath,amssymb,
 aps,superscriptaddress,]{revtex4-2}

\usepackage[utf8]{inputenc}
\usepackage[english]{babel}
\usepackage[a4paper,margin=2.0cm]{geometry}


\usepackage{amsmath,amsfonts,amssymb,dsfont,amsthm}
\usepackage{indentfirst}
\usepackage{graphicx}
\usepackage{booktabs}
\usepackage{siunitx}
\usepackage{lipsum}
\usepackage{listings}
\usepackage{tasks}
\usepackage{amssymb}
\usepackage[T1]{fontenc}

\usepackage{comment}
\usepackage{mathtools}
\usepackage{extpfeil}
\usepackage{xcolor}
\usepackage{hyperref}

\usepackage{url}
\usepackage{natbib}
\usepackage{makecell}

\usepackage{physics}
\usepackage{graphicx}
\usepackage{dcolumn}
\usepackage{bm}
\usepackage{tikz}
\usepackage{pgf}
\usepackage{tikzscale}
\usetikzlibrary{angles,calc,intersections,quotes,arrows.meta, shapes,decorations.markings, external}

\newcommand{\ie}{\textit{i.e.}}

\newcommand{\ba}[1]{\begin{align} #1 \end{align}}

\begin{document}

\title{Communication-constrained nonlocal correlations}

\author{Lucas Pollyceno}
\email{lucas.da.silva.pollyceno@ug.edu.pl}
\affiliation{\mbox{International Centre for Theory of Quantum Technologies, University of Gda{\'n}sk, 80-308 Gda\'nsk, Poland}}

\author{Denis Freudenheim}
\affiliation{Instituto de Física Gleb Wataghin, Universidade Estadual de Campinas, 13083-859, Campinas, Brazil}

\author{José Nogueira}
\affiliation{Instituto de Física Gleb Wataghin, Universidade Estadual de Campinas, 13083-859, Campinas, Brazil}

\author{Anubhav Chaturvedi}
\affiliation{\mbox{Division of Quantum Optics and Information, Institute of Theoretical Physics and Astrophysics,} \mbox{Faculty of Mathematics, Physics and Informatics, University of Gda{\'n}sk, 80-308 Gda\'nsk, Poland}}

\author{Rafael Rabelo}
\affiliation{Instituto de Física Gleb Wataghin, Universidade Estadual de Campinas, 13083-859, Campinas, Brazil}

\author{Marcin Paw{\l}owski}
\affiliation{\mbox{International Centre for Theory of Quantum Technologies, University of Gda{\'n}sk, 80-308 Gda\'nsk, Poland}}

\begin{abstract}

Identifying the physical grounds distinguishing quantum theory from broader probabilistic frameworks remains an open challenge. Communication-based proposals -- most notably the principles of \textit{impossibility of superluminal signaling} and \textit{information causality} (IC) -- highlight the role of communication in ruling out ‘unphysical' theories and offer an operational rationale on why quantum predictions prevail over these alternative models. Nevertheless, most such developments rely on communicating parts optimizing over specific tasks, such as \textit{communication complexity problems} and \textit{random access codes} (RAC). In this work, we systematically extend this communication-based approach. We characterize the class of communication tasks relevant for this context, and employ the general information-theoretic framework to derive new operational constraints preventing such ‘unphysical' behaviors. Remarkably, our results reveal a broad family of previously undetected implausible behaviors, independent of any particular encoding or decoding strategy, reinforcing the role of communication as a fundamental lens through which physically meaningful theories can be identified.
    
\end{abstract}
\maketitle

\section{Introduction}

Quantum mechanics stands as one of the most successful scientific theories, achieving unprecedented agreement between experimental observations and theoretical predictions. Nevertheless, it remains unclear whether quantum mechanics is a fundamental theory of nature -- reconcilable with other physical theories such as general relativity -- or merely a predictive model confined to a particular regime of application. One major challenge in this regard is that quantum mechanics still lacks clear axiomatic physical statements, remaining largely an abstract description of Hilbert space formalism. Some attempts address this issue by starting from outstanding features of quantum mechanics -- such as the no-cloning theorem \cite{Wootters1982ASQ}, or teleportation \cite{teleport} -- as primitive physical notions, from which the quantum formalism can emerge \cite{PhysRevA.84.012311, PhysRevA.81.062348}. However, it remains open whether such remarkable ‘quantum features' uniquely identify quantum theory, since many of these properties are not exclusive to quantum mechanics but can also emerge in broader probabilistic frameworks \cite{GPT}.

Other approaches start from different physical grounds, building up a set of reasonable axioms that should hold for any physical theory \cite{PRbox, Gisin_2020, NT, IC, ML, OL}. For instance, regardless of the mathematical formalism, it is widely accepted that any reasonable physical theory should agree with \textit{the impossibility of super-luminal signaling} (the nonsignaling principle). Interestingly, however, while nonsignaling (NS) theories allow for several of the phenomena considered quantum -- such as no-cloning \cite{nocloning, nobroadcasting}, teleportation \cite{teleportationGPT}, entanglement swapping \cite{SwappingNS} -- NS alone fails to distinguish quantum theory from other probabilistic theories \cite{PRbox}. Consequently, it becomes crucial to identify the physical grounds underlying quantum theory that distinguish it from the other NS theories.

Although NS encompasses many of these ‘quantum features', it has been shown to permit implausible consequences \cite{vandam2005implausible, IC}, such as the trivialization of distributed computing problems \cite{vandam2005implausible}. This fact has motivated the formulation of additional physical principles aiming to single out quantum theory and prevent such unphysical behaviors, notably the \textit{non-trivialization of communication complexity problems} (NTCC) \cite{NT} and the \textit{information causality} principle (IC) \cite{IC}. Of particular relevance, any NS correlation exceeding the maximum quantum violation of the Clauser-Horne-Shimony-Holt (CHSH) inequality \cite{CHSH} (the well-known Tsirelson's bound \cite{tsirelson}) leads to a violation of IC \cite{IC}, building an alternative proof for the boundary predicted by the quantum theory formalism based solely on information-theoretic grounds. However, it remains unclear whether IC fully distinguishes quantum mechanics from all other probabilistic theories \cite{Allcock_2009, Yang_2012, ICnoisy, multiIC}.

The distinctive character of these proposals is their emphasis on the fundamental role of \textit{communication} for the understanding of physical theories. In these cases, the ‘unphysical' feature of NS resources is identified by the implausible outstanding performance of communicating parties on tasks such as \textit{communication complexity} (CC) \cite{CC, CCbook} or in \textit{random access codes} (RAC) \cite{conjcoding, QRAC}. Given the success of this framework, a fundamental question emerges of whether other forms of communication tasks could also reveal some sort of implausible consequences, thereby enhancing the current description of the set of reasonable physical theories. In fact, it seems unlikely that only one specific task, such as the RAC, could entirely identify the extent of physically meaningful theories. In line with that, slight modifications of the RAC task have already demonstrated usefulness in particular contexts, such as multipartite scenarios \cite{multiIC} and cryptographic security proofs \cite{SecurityIC}.

In this work, we systematically investigate this question and identify the extent to which the \textit{communication-based} approach can reveal such implausibilities in NS theories. In particular, we introduce \textit{general communication scenarios} as a systematic framework for identifying and characterizing the relevant communication tasks. Notably, this notion allows us to overcome both the inherent protocol-dependence and task-dependence present in all previous \textit{communication-based} formulations \cite{IC, NT}. We then show that these results establish novel operational bounds on physically meaningful theories within the geometric information-theoretic framework \cite{Yang_2012, Chaves2, EntropicQuantum}. This allows us to uncover a broad range of scenarios that reveal implausible consequences -- cases that previous approaches as IC and NTCC failed to detect.

\section{Preliminaries}

Originally, the IC principle is intrinsically formulated by means of the $(n\mapsto m)$ \emph{random access code} (RAC) task, wherein the parties utilize nonlocal correlations assisted by a classical communication channel of bounded capacity $C$. In this case, Alice has initially a randomly sampled bit-string $X\equiv(X_0,\ldots, X_{n-1})\in\{0,1\}^n$ of length $n$, and encodes it in a message to be sent to Bob, the receiving party. Bob, in turn, has to produce a guess $B_i$, aiming to optimize the success guessing probability,
\ba{\label{eq:rac_suc}p_{\text{s}} = \frac{1}{n2^n} \sum_i^{n-1}\sum_{\mathbf{x}} p(x_i = b_i |i, \mathbf{x}).}
Within this context, IC states that
\textit{the total potential information Bob can gain about $X$ cannot exceed the capacity $C$ of the classical communication channel}. Different operational formulations of the IC statement have been proposed in distinct contexts \cite{IC, Chaves2, ICnoisy, multiIC, SecurityIC}. For this work, we restrict our attention to the currently most stringent formulation for bipartite communication scenarios, as introduced in \cite{ICnoisy}:
\ba{\label{eq:noisy_original} \sum_{i=0}^{n-1} I(X_i : B_i) \le C.}
where $I(X_i:B_i)$ denotes Shannon's mutual information.

Quantum theory satisfies IC, while there are correlations stronger than those allowed by quantum mechanics that violate IC and are therefore ruled out by the principle \cite{IC}. This becomes evident when considering the simplest noiseless $2\mapsto 1$ case ($C=1$), where parties have access to a NS correlation from a Bell scenario $p_{\text{Bell}}(a,b|x,y)$ --- $a,b,x,y \in \{0,1\}$ --- fulfilling $a \oplus b = xy$ \cite{PRbox}. In this case, Alice inputs $x=x_0\oplus x_1$ and sends the message $m=a\oplus x_1$, while Bob outputs $b_y = m\oplus b$ -- the so-called van Dam protocol \cite{vandam2005implausible}. It is straightforward to check that it yields $p_{\text{s}} = 1$, while the maximum quantum performance is given by Tsirelson's bound: $p_{\text{s}} \le p_Q \approx 0.854$ \cite{EARAC, tsirelson}. At the same time, such a procedure clearly violates IC, since $I(X_0:B_0)=I(X_1:B_1)=C=1$, thereby being ruled out by criterion \eqref{eq:noisy_original}. Of particular relevance is the fact that any nonlocal correlation $p(a,b|x,y)$ exceeding the quantum performance bound $p_Q$ is ruled out by IC, for some $C\in [0,1]$ \cite{IC, ICnoisy}. 

Despite IC partially explaining the boundary of quantum nonlocality, it remains open whether IC fully recovers the set of quantum correlations. Indeed, there are post-quantum correlations that do not achieve the quantum bound $p_Q$ for which the violation of inequality \eqref{eq:noisy_original} is unclear \cite{Allcock_2009, requiresMultipartite}. This fact motivated the search for more general criteria describing the IC notion by employing the systematic information-geometric framework \cite{Chaves2}.

\subsection{Information-theoretic relations}\label{Qcausalstructure}

Consider a general set of random variables $S=\{X_0, X_1, ..., X_{n-1}\}$, possibly including both classical and quantum variables. We say that the variables in $S$ are constrained by the rules of information theory if they necessarily satisfy a finite set of entropic relations, commonly known as the \textit{basic inequalities} \cite{Yeung, EntropicQuantum}. The entire set of informational inequalities is sufficiently expressed in terms of two elemental inequalities \footnote{Meaning that any entropic inequality expressing information-theoretic constraints on the random variables S arises from suitable combinations of these elemental inequalities.}: \textit{(i) strong subadditivity}: $H(A, C) + H(B, C) \geq H(A, B, C) + H(C)$; and \textit{(ii) weak monotonicity}: $H(A, B) + H(A, C) \geq H(B) + H(C)$. Of particular relevance is the fact that obtaining all elemental inequalities for $S$ requires prior knowledge of the relations between its variables, as some quantum variables cannot be jointly described. This is evident from the fact that some quantum measurements leave no well-defined post-measurement state, preventing any joint description of the measured variable and the system afterward. This fact introduces the necessity of defining the concept of \textit{coexisting sets} -- \ie, the subsets of $S$ for which all contained random variables admit a joint description \cite{Weilenmann_2017}. In this case, the set $S$ is limited by the elemental inequalities when \textit{(i) and (ii)} hold for each of the coexisting sets of $S$.

After incorporating the elemental inequalities associated with $S$, a further refinement of the information-theoretic description can be achieved by considering the causal relations within $S$. Formally, these relations are represented by a \textit{directed acyclic graph} (DAG), where each node corresponds to one variable in $S$ (see, for instance, Fig.~\ref{fig:noiseless}) and each directed edge represents a causal relation between the variables associated with the connected nodes. In this representation, the value of each variable $X_i$ is influenced only by its \emph{causal parents} -- i.e., variables $X_{j}$ such that $X_{j} \rightarrow X_{i}$ is a directed node in the DAG. These causal dependencies can be expressed in information-theoretic terms through two key types of constraints: \textit{(iii) conditional independence relations} (CIR), such as $I(A: C|B) = 0$ (for $A \rightarrow B \rightarrow C$); and \textit{(iv) data-processing inequalities}, such as $I(A: B') \le I(A: B)$, for $B \rightarrow B'$. By combining all constraints from (i) to (iv), we essentially encompass all assumptions that come from information theory and the causal framework.

The last step consists of marginalizing the obtained information-theoretic description to the set of jointly observable variables $S_{M}$. In fact, even though the coexisting sets were considered from the outset, $S$ may still contain latent variables that causally influence the statistics but are not directly accessible from the observed data \cite{Instrumental}. Thus, we have to eliminate from the description every term not present in $S_{M}$, \ie, the unobserved variables. This is achieved through standard variable elimination methods, such as Fourier-Motzkin elimination \cite{FM}. After this procedure, we have a set of inequalities encompassing all possible entropically formulated principles, built with the helm of quantum information and causality theories. 

As originally demonstrated in Ref.~\cite{Chaves2}, this method recovers the standard IC criterion, with inequality \eqref{eq:noisy_original} corresponding to one specific inequality among several others. In fact, this approach enables the derivation of 54 new informational inequalities, some of which impose stricter bounds than those obtained from the original formulation in \eqref{eq:noisy_original}. Nevertheless, due to the \textit{inherent protocol dependence}, it remains an open question whether all post-quantum correlations can be excluded by the IC principle. Indeed, so far, all IC analyses have assumed that Alice and Bob follow the van Dam protocol. Although this protocol achieves optimal performance when the parties share PR boxes, it does not necessarily hold optimal for other types of correlations. 

Beyond the protocol dependence of current formulations, most IC analyses restrict the parties to optimizing their performance in the specific task of RAC~\eqref{eq:rac_suc}. However, as previously discussed in the preamble, considering alternative tasks has become essential, since recent developments have proven useful for extending the analysis to multipartite scenarios~\cite{multiIC}, enabling the recovery of the monogamy of nonlocality and ensuring cryptographic security~\cite{SecurityIC}. These insights motivate the framework introduced in the next section, where we identify the communication tasks that should be analyzed within this context.

\section{General communication scenarios}

\begin{figure}[t!]
    \centering
    \includegraphics[width=1\linewidth]{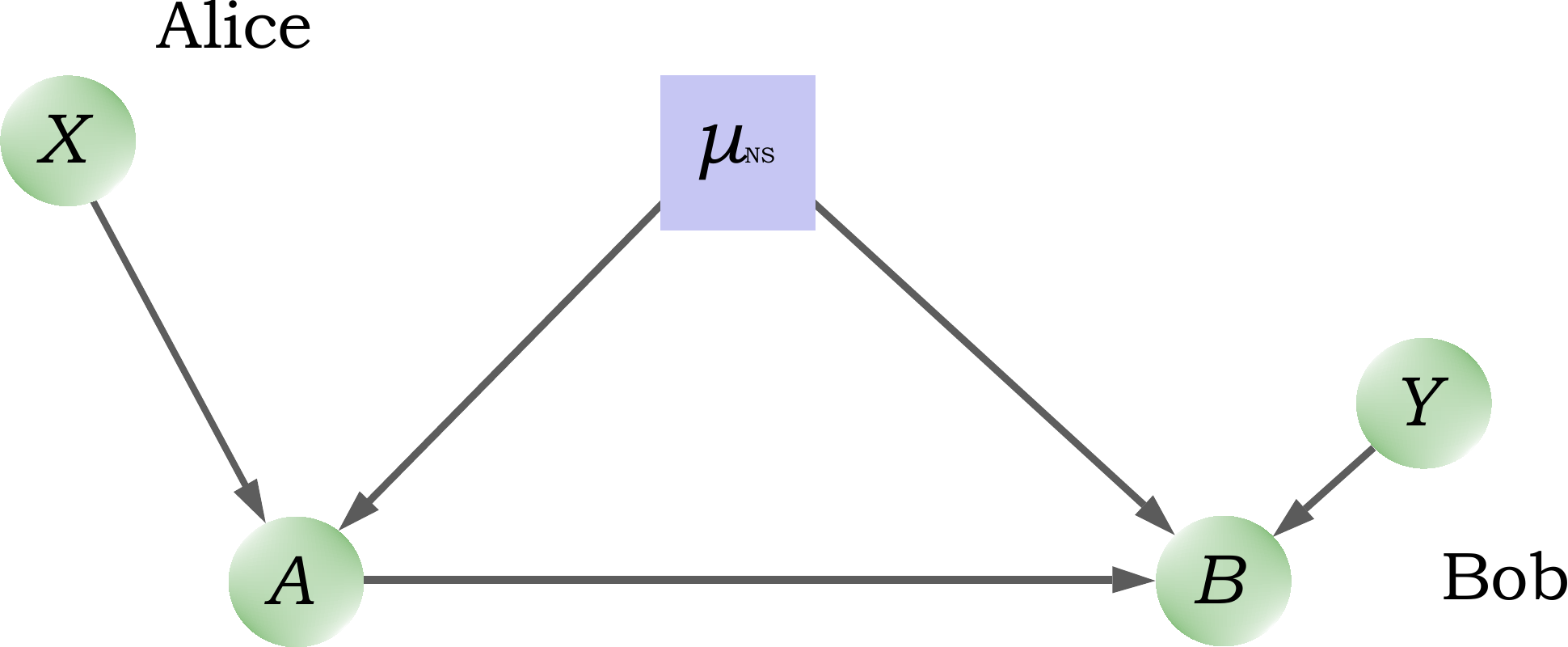}
    \caption{Causal structure associated with general communication scenarios. The parties share a no-signaling (NS) resource $\mu_{\text{NS}}$, which assists Alice in preparing a classical message $A$ that encodes the initial data $X$. Upon receiving the message and supported by the shared resource, Bob produces an outcome $B$ based on a randomly selected input $Y$.}
    \label{fig:noiseless}
\end{figure}

A general communication (GC) scenario can be described in terms of the so-called \textit{prepare-and-measure} (PM) framework. In this setting, one party, Alice, possesses a dataset denoted by $X$ and encodes it into a message by preparing a physical system, whose state we denote by $A$, which is then sent to a second party, Bob. The receiver performs a measurement, labelled by $Y$, on the received system and produces an output $B$. More generally, and of particular relevance to the present work, the parties may also exploit pre-established correlations, which we denote by $\mu_{NS}$. A communication scenario is then fully specified by the tuple $(|X|,|Y|;|A|,|B|)$ and is illustrated in Fig.~\ref{fig:noiseless}. Naturally, the parties may employ different types of physical resources to encode their messages. However, since we aim to discriminate among the correlating resources, we restrict our discussion to classical communication. Within this context, the most general description of such an experiment is given by a collection of probability distributions $p(a,b|x,y)$, often referred to as the \textit{behavior} of the experiment. Note that, unlike in standard PM scenarios, here the message $A$ is treated as an observed variable. Indeed, since $A$ is a classical variable, it can be observed without interference \cite{wiringbell, rout2025}. Throughout this paper, we denote random variables by capital letters such as $X$ and $A$, and use the shorthand notation $p(x) = p(X = x)$ to denote the probability that random variable $X$ takes the value $x$.

The probabilities corresponding to the experiment admit a quantum description if they can be written as:
\begin{align}\label{eq:quantum_model}
p(a,b|x,y) = Tr\left[\rho_{AB}(M_{a|x}\otimes M_{b|a,y})\right],
\end{align}
which represents the probability of obtaining outcomes $a$ and $b$ given that the parties share a composite quantum state $\rho_{AB} \in \mathcal{L}(\mathbb{C}^d\otimes \mathbb{C}^d)$ and perform quantum measurements described by POVMs $\{M_{a|x}\}_{a}$ and $\{M_{b|a,y}\}_{b}$. Here,  the parties use the most general adaptive strategy \cite{Adaptive, Vieira_2023}, where Alice communicates her outcome $a$, which is then used by Bob to adapt his measurement choice.

Notice the equivalence of definition \eqref{eq:quantum_model} with the one from the context of Bell scenarios \cite{Bellnonlocality}. In fact, from a geometric perspective, the probability distributions observed in GC scenarios correspond to Bell-scenario correlations where some components are effectively ignored \cite{wiringbell, rout2025}, specifically,
\ba{\label{eq:behavior_ns}
p(a,b|x,y) = \sum_{y'} p_{\mathrm{Bell}}(a,b|x,(y, y'))\delta_{y'= a}.}
This naturally defines the probability distributions in GC scenarios that can be achieved with resources constrained solely by the nonsignaling (NS) principle, where $p_{\mathrm{Bell}}(a,b|x,(y, y'))$ describes a Bell scenario correlation that respects the NS conditions: 
\begin{subequations}
\ba{
\sum_a p_{\mathrm{Bell}}(a,b|x,(y, y')) = p_{\mathrm{Bell}}(b|(y, y')) \; \forall\; x, y, y', b \\
\sum_b p_{\mathrm{Bell}}(a,b|x,(y, y')) = p_{\mathrm{Bell}}(a|x) \forall \;x,y,y',a.
}
\end{subequations}
Given a GC scenario, the inputs and outputs of the associated Bell scenario are fixed (and coincide with those of the former), except for the additional input $Y'$ on Bob's side, whose cardinality matches that of the classical message A. Interestingly, differently from previous approaches \footnote{In fact, without this link, we may always care about finding suitable encoding and decoding functions, which in principle there might be infinitely many ways of doing, even for small scenarios} within this framework, we efficiently characterize all correlations that NS resources can produce in a GC scenario, as determined by the tuple$(|X|,|Y|;|A|,|B|)$. Moreover, from \eqref{eq:behavior_ns} it follows that the set defined by the NS condition in GC forms a polytope, which can be efficiently characterized in terms of its vertices using \textit{vertex enumeration} methods \cite{panda}. 

Not surprisingly, as in Bell scenarios, GC correlations defined by \eqref{eq:behavior_ns} may not be described by a quantum model, as defined in equation \eqref{eq:quantum_model}, in which case they are commonly referred to as \textit{post-quantum}. This feature can be identified through \textit{nonclassicality witnesses}. Specifically, when parties are restricted to classical resources (\ie, $\mu_{\text{NS}} \longrightarrow \lambda$, where $\lambda$ denotes any classical shared resource), the  probability distributions are described in terms of the classical model:
\ba{\label{eq:classical}
p(a,b|x,y) = \sum_{\lambda} p(\lambda) p(a|x, \lambda) p(b|a,y,\lambda).}
Note that this model differs from the standard \textit{local hidden-variable} model, since Bob’s outcome may depend on Alice’s message. Equation~\eqref{eq:classical} defines the set of \emph{classical correlations} compatible with the given GC scenario. This set is convex and, more specifically, forms a polytope whose extremal points correspond to deterministic strategies for Alice and Bob. As a consequence, classicality in GC scenarios can equivalently be characterized through an inequality-based formulation, given by constraints of the form
\begin{align}\label{eq:Bell-type}
    \sum_{a,b,x,y} C_{ab}^{xy} \:p(a,b|x,y) \le \beta_C,
\end{align}
where the coefficients $C_{ab}^{xy}$ and the bound $\beta_C \in \mathbb{R}$ define a particular inequality. That is, a correlation admits the classical model \eqref{eq:classical} if and only if it satisfies all inequalities of the form \eqref{eq:Bell-type} for the given GC scenario. Naturally, quantum and nonsignaling correlations as described by equations \eqref{eq:quantum_model} and \eqref{eq:behavior_ns} may present nonclassicality, revealed by their ability to outperform the classical bound $\beta_C$, achieving, respectively, quantum and nonsignaling bounds $\beta_Q$ and $\beta_{\text{NS}}$, where we observe the hierarchy $\beta_C \le \beta_Q \le \beta_{\text{NS}}$. These \textit{nonclassicality witnesses}, inequalities \eqref{eq:Bell-type}, in turn, capture all relevant communication tasks in a given GC scenario\footnote{Evidently, they do not cover every possible task, but they constitute the optimal ones for witnessing the separation between quantum and nonsignaling nonlocality.}. For instance, the RAC task can be identified within this framework, corresponding to a specific nonclassicality witness \eqref{eq:Bell-type} in standard PM scenarios \cite{SDI}.

The simplest GC scenario is defined by $|X| = 3$, $|Y| = 2$, $|B| = 2$, and $|A| = 2$. By employing \textit{facet enumeration} methods \cite{panda}, we identify 576 Bell-type inequalities of the form \eqref{eq:Bell-type}, which can be grouped into 12 equivalence classes under relabeling symmetries. A detailed account of this characterization is provided in the Supplementary Material \ref{sec:characterization}, where the representatives of the 12 classes are explicitly listed. Table \ref{table:1} presents the corresponding bounds for each class. Upper bounds on quantum values were obtained using the NPA hierarchy \cite{NPA}, while the nonsignaling bounds were derived by enumerating the vertices of the nonsignaling polytope associated with the respective $(3,4;2,2)$ Bell scenario \cite{extremals}. We remark that, although some tasks reveal no separation among the different sets, most of the witnesses show clear differences in performance across the considered resources. Furthermore, it is important to stress that, due to the nature of the framework, the presented bounds are protocol-independent, overcoming a common limitation of previous approaches.

\begin{table}[t!]
  \centering
  \caption{Table containing the bounds regarding the twelve task of the $(3,2;2,2)$ GC scenario for the different kind of resources: classical (C), quantum (Q) and non-signaling (NS). The bounds from the impossibility of CC problems become trivial \cite{NT} (NTCC) and the information-theoretic bounds from inequality~\eqref{eq:relevant} are reported in the last columns.}
  \label{table:1}
  \setlength{\tabcolsep}{8pt}   
  \renewcommand{\arraystretch}{1.15} 

  \begin{tabular}{c |c c c c c}
    \toprule
    Ineq. & $C$ & $Q$ & $NS$ &NTCC& Eq.~\eqref{eq:relevant} \\
    \midrule
    1  & 0.0 & 0.0    & 0.0 & 0.0 &  0.0 \\
    2  & 1.0 & 1.0    & 1.0 & 1.0 &  1.0 \\
    3  & 1.0 & 1.207 & 1.5 & 1.5 &  1.5 \\
    4  & 1.0 & 1.207 & 1.5 & 1.5 &  1.5 \\
    5  & 1.0 & 1.207 & 1.5 & 1.316 &  1.5 \\
    6  & 1.0 & 1.207 & 1.5 & 1.5 &  1.5 \\
    7  & 1.0 & 1.207 & 1.5 & 1.5 &  1.5 \\
    8 & 1.0 & 1.207 & 1.5 & 1.316 &  1.5 \\
    9 & 2.0 & 2.25 & 3.0 & 2.816 &  2.723 \\
    10 & 2.0 & 2.25 & 3.0 & 2.816 &  2.723 \\
    11  & 2.5 & 2.65 & 2.75 & 2.75 &  2.658 \\
    12  & 2.5 & 2.65 & 2.75 & 2.75 &  2.658 \\
    \bottomrule
  \end{tabular}
\end{table}

One important feature of these newly derived tasks is their inequivalence to the standard RAC and CC tasks. This distinction becomes particularly evident when examining their explicit algebraic form. A complete characterization of all identified tasks is provided in the Supplementary Material; here, we highlight two representative examples, corresponding to tasks 11 and 12 in Table~\ref{table:1}:
\begin{subequations}\label{eq:task7}
\begin{align}
 p((a\oplus &1)b = ay\oplus 1 \mid x=0)
 + p(by =0 \mid x=1) \nonumber\\
 &
 + p(b(a\oplus y \oplus 1) = ay \mid x=2)
 \le 5/2 ,\\[1.0em]
 p((a\oplus &1)b = ay\oplus 1 \mid x=0)
 + p(ab =by \mid x=1)\nonumber \\
 &
 + p(a(y\oplus1) =b(y\oplus 1) \mid x=2)
 \le 5/2 .
\end{align}
\end{subequations}
In both cases, a clear separation between the quantum and NS bounds is observed, while these tasks admit no natural interpretation in terms of RAC protocols that could account for such a separation. Moreover, although CC encompasses a broad class of communication tasks, following the methods present in Ref.~\cite{NT}, we demonstrate in Table~\ref{table:1} that for~\eqref{eq:task7} and several other identified tasks, the impossibility of trivializing CC fails to reveal any implausible consequences. Interestingly, however, in the next chapter we demonstrate that the advantage over quantum resources in the identified GC tasks may also lead to implausible consequences that remain undetected by existing communication-based frameworks. To address this, we present a systematic analysis of the information-theoretic constraints governing GC scenarios.

\section{Information-theoretic description of GC scenarios}
\begin{figure}[t!]
    \centering
    \includegraphics[width=1\linewidth]{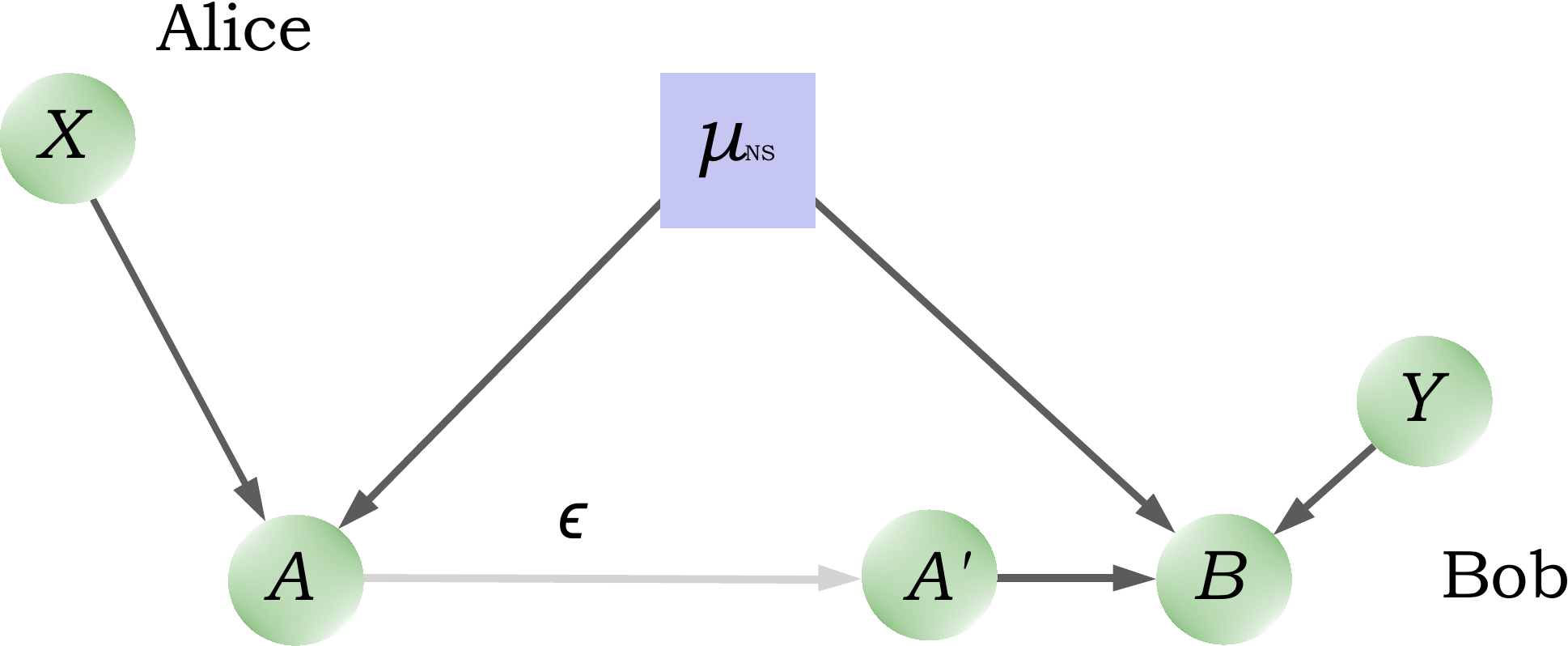}
    \caption{Causal structure associated with general communication scenarios depicted in Fig.~\ref{fig:noiseless}, extended to include classical noisy communication specified by $\epsilon$, yielding the effective message $A'$ received by Bob.}
    \label{fig:dag}
\end{figure}

As aforementioned in the preliminaries, the IC principle emerges as a natural constraint on the amount of knowledge that communicating parties can gain. This notion is formulated more generally in terms of the causal relation of the communication scenario, as illustrated in Fig.~\ref{fig:dag}. Here, Alice holds an initial dataset $X$ and encodes it into a message $A$, making use of her share of a pre-established correlating resource $\mu_{\text{NS}}$. Bob receives the message through a noisy classical communication channel, so that the effective message at his end is denoted by $A'$, with the noise level governed by the parameter $\epsilon$. Supported by $A'$ and all available local resources, Bob produces an output $B_y$, where the input $Y$ specifies which information about $X$ he attempts to access in that round. It is important to emphasize that, unlike previous approaches, this description captures the minimal set of concepts related to the notion introduced within the IC principle. 

Up to this point, we have not assumed any particular structure for $X$, nor imposed any specific relation between the sent and received messages. Likewise, in contrast to RACs, the input $Y$ is not required to correspond to a particular guess made by Bob. The only essential constraint here is that $|Y|\ge 2$; otherwise, the scenario would provide Bob with no potential choice, in which case the IC principle plays no role. Thus, Fig.~\ref{fig:dag} depicts the minimal causal structure underlying the IC notion. Within this structure, the proposed information-geometric framework allows us to extract all informational constraints implied by the causal relations. In this case, the quantum description is obtained considering the NS resource being reduced to a quantum state -- \ie, $\mu_{\text{NS}} \longrightarrow\rho_{AB}$. In the simplest case with $|Y|=2$, the only relevant CIR are $I(X:\rho_{AB}) =0$ and $I(A':X, \rho_{B} | A) = 0$. The corresponding marginal scenario is then $\{\{X, A, A', B_y\}_y\}$, where $\rho_{B}$ denotes Bob’s local part of the correlating resource $\rho_{AB}$. Applying FM elimination, the resulting Shannon cone yields all possible constraints associated with the causal structure in Fig~\ref{fig:dag} (see the Supplementary Material for details). In this case, only two nontrivial inequalities arise, both of the form (indexed by $y\in\{0,1\}$):
\begin{multline}\label{eq:relevant}
I(X:A, A', B_y) + I(A: A', B_{y\oplus 1}) \le\\
H(A) + I(A:A').
\end{multline}

\noindent 
In fact, these are the only inequalities forbidding NS correlations in the analyzed GC scenarios. Inequality \eqref{eq:relevant} naturally recovers the essence of IC: the left-hand side represents the spread of knowledge about the initial data $X$ and the encoded message $A$ through the causal structure to Bob, while the right-hand side limits this by the information encoded in $A$ and the amount effectively received in $A'$. In the noiseless case, inequality \eqref{eq:relevant} reduces to the \textit{instrumental scenario} inequality \cite{Instrumental}, $I(X: A, B_y)\le H(A)$, which does not capture the same notion as IC since it neglects Bob’s potential choice. Interestingly, inequality \eqref{eq:relevant} makes explicit that the information-theoretic description of the DAG in Fig.~\ref{fig:dag} constrains Bob’s accessible knowledge of the sources of uncertainty in the causal structure -- namely, the initial data $X$ and the noisy channel governed by $\epsilon$. As we show in the following sections, incorporating additional sources of uncertainty naturally leads to stronger IC-related criteria.

Table \ref{table:1} reports the bounds implied by inequality \eqref{eq:relevant} for the different communication tasks in the simplest GC scenario ($3,2;2,2$). These results are obtained following the methods described in the Supplementary Material. In this setting, the parties communicate through a binary symmetric noisy channel, which flips the message bit $A$ with probability $p(a'=a\oplus 1|a)=\epsilon \in [0,1/2)$ for $A\in\{0,1\}$. The optimal results are achieved in the limit $\epsilon \longrightarrow 1/2$. In this case, we observe that GC tasks of the form~\eqref{eq:Bell-type} can indeed lead to implausible consequences when assisted by NS resources, such as violations of the information-theoretic bound given by inequality~\eqref{eq:relevant}. Of particular relevance, this occurs for the tasks identified by representatives 9, 10, 11 and 12. Notably, for classes 11 and 12 (see~\eqref{eq:task7}), no implausible consequences are detected by any of the previously proposed approaches. The findings obtained for classes 3, 4, 6, and 7, on the other hand, suggest the communication-based framework is insufficient to fully characterize the quantum boundary in GC scenarios, as no violations are observed in these cases. Importantly, due to the structure of the framework, the derived bounds are protocol independent and, unlike previous approaches, cannot be improved through more suitable protocols.

One important feature of the new criteria becomes clear when analyzing inequality \eqref{eq:relevant} in the context of the PR-box correlation, as in Ref.~\cite{IC}. In particular, notice that the standard protocol for the $2\mapsto1$ RAC described in the preliminaries does not directly violate inequality \eqref{eq:relevant}, and a modified protocol must be considered instead. Following the same ideas introduced for GC scenarios, we let Alice have a binary input $X$, encode the message $x\oplus a$ into the binary symmetric communication channel, and let Bob produce the output $a'\oplus b$. This yields final outputs for Bob such that $p(b_0 = x) = 1-\epsilon$ and $p(b_1 = 0) = 1-\epsilon$, which saturates the left-hand side of inequality \eqref{eq:relevant}. Interestingly, in this case, Bob can perfectly certify the channel behavior for $y=1$, regardless of the noise parameter $\epsilon$. Indeed, when considering $\epsilon \longrightarrow 1/2$ and the NS correlation specified as $p_{\text{Bell}}(a,b|x,y) = \gamma\delta_{a\oplus b,xy}/2 + (1-\gamma)/4$, criterion \eqref{eq:relevant} is violated for all $\gamma \gtrsim 0.78$. This, nevertheless, leaves a gap with respect to the quantum bound $\gamma_Q \lesssim 0.707$. Still, this result demonstrates consistency, since the derived bound coincides with the one implied by the original criterion in the particular case of the $2\mapsto1$ RAC with noiseless communication \cite{IC}. This equivalence can be understood by noticing that both cases involve two sources of uncertainty. Consequently, in the context of bounding NS correlations using the IC notion, considering two initial random bits and one perfect channel is equivalent to having one random variable and one noisy binary channel. Naturally, this fact motivates investigating the inclusion of additional sources of uncertainty in the IC causal structure (Fig.~\ref{fig:dag}) in the context of GC scenarios.

\subsection{Adding sources of uncertainty}

\begin{figure}[t!]
    \centering
    \includegraphics[width=1\linewidth]{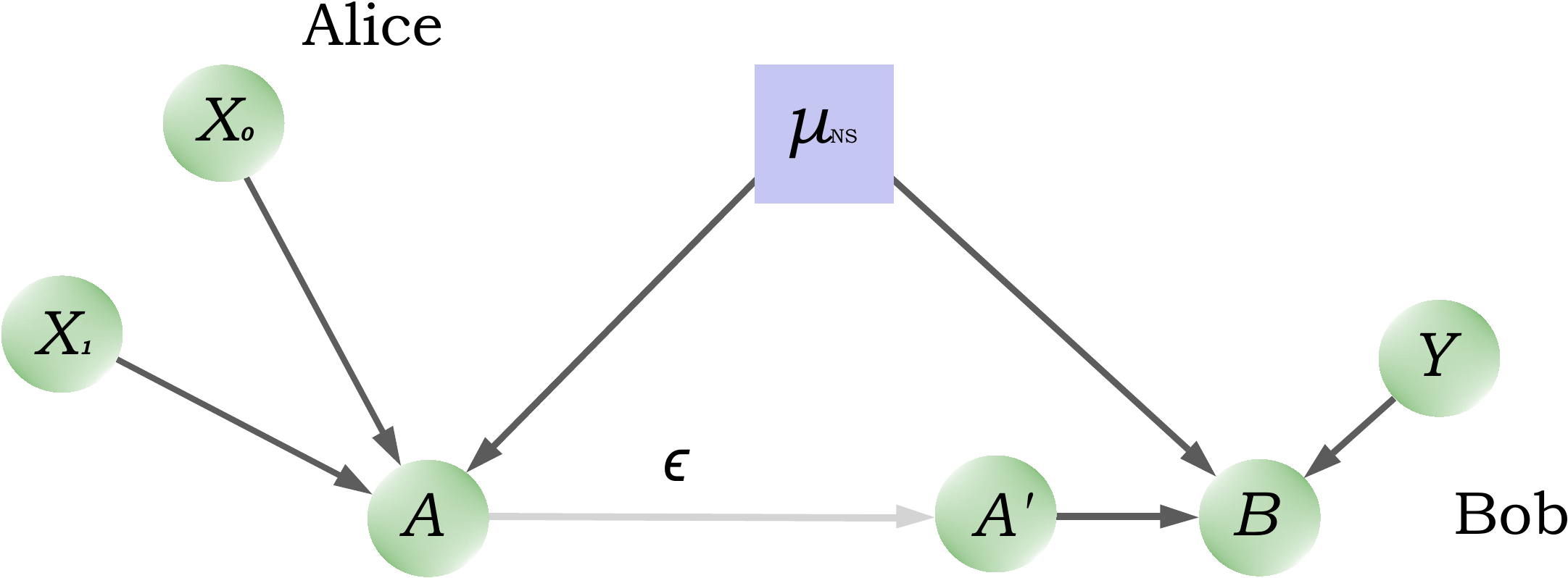}
    \caption{Causal structure considering three sources of uncertainty: $X_0, X_1$ and the classical communicating channel specified by $\epsilon$.}
    \label{fig:dag2}
\end{figure}

Fig.~\ref{fig:dag2} describes the case where Alice initially holds two independent sets of data, $X_0$ and $X_1$. The only relevant CIR in this scenario are $I(X_0:X_1)=0$, $I(X_0X_1:\rho_{AB})=0$, and $I(A':X_0X_1 \rho_{B}|A)=0$, with the marginal scenario given by $\{\{X_0, X_1, A, A', B_y\}_y\}$. Despite this seemingly small change, performing Fourier-Motzkin elimination becomes highly challenging. In fact, the number of inequalities grows double-exponentially with each additional variable in the causal structure, rendering the direct FM elimination method computationally impractical. Nonetheless, in the Supplementary Material, we explain how to overcome this limitation by applying FM to a significantly reduced set of inequalities. In generic terms, the method is intuitively understood considering that our main focus is to identify implausible consequences of outperforming quantum theory boundaries in the context of GC tasks (such as those in Table \ref{table:1}). Because of the generality of the method, however, many constraints in Shannon's cone description play no role in bounding the performance of the relevant tasks. In Supplementary Material \ref{ap:selection}, we detail how to efficiently isolate the constraints that effectively bound a GC correlation $p(a,b|x,y)$ of interest. As a result, we obtain a much smaller set of inequalities, allowing FM to be carried out efficiently.

One crucial step in this case is identifying the post-quantum correlations in the GC scenario of interest. The simplest GC scenario compatible with the causal structure in Fig.~\ref{fig:dag2} is specified by $|X| = 4$, $|Y| = 2$, $|B| = 2$, and $|A| = 2$. According to definition \eqref{eq:behavior_ns}, the relevant post-quantum correlations here are given by the NS vertices of the associated Bell scenario $(4,4;2,2)$. In total, there are $194340$ vertices, grouped into $20$ equivalence classes (the complete characterization is provided in \cite{extremals}), each potentially leading to different entropic criteria. Following the methods detailed in Supplementary Material \ref{ap:entropic_characterization} we obtained all the
optimal information-theoretic criteria identifying the post-quantumness of all 20 classes of extremals. Interestingly, however, all obtained inequalities can be shortly summarized in the form of the following four inequalities:
\ba{\label{eq:newic1}I(X_j : X_{j\oplus1}, A'B_l)& + I(X_{j\oplus1} : A'B_{l\oplus1}) \le\\ &I(A:A') + I(X_{l}:X_{l\oplus1}),\nonumber}
where $j,l\in\{0,1\}$. Table \ref{table:noise_resilience} presents the explicit instances of Eq.~\eqref{eq:newic1} that are optimal for each of the 20 classes of extremal NS correlations. These newly derived criteria demonstrate the internal consistency of the framework, as they provide a strictly more accurate operational form for the IC notion. This becomes evident by observing that Eq.~\eqref{eq:newic1} reduces to the previous formulation (Eq.~\eqref{eq:noisy_original} for $n=2$) upon considering the relaxations: $[I(X_j : X_{j\oplus1}, A'B_l) \ge I(X_j :B_l)$, $I(X_{j\oplus1} : A'B_{l\oplus1}) \ge I(X_{j\oplus1} : B_{l\oplus1})$, for $j=l=0]$. Indeed, the original formulation of Eq.~\eqref{eq:noisy_original} implicitly assumes a restricted marginal scenario of the form $\{\{ X_0,B_0\},\{X_1,B_1\}, \{A,A'\}\}$, where part of the information available to Bob is effectively discarded. Within this perspective, the generality of the criteria in Eq.~\eqref{eq:newic1} becomes justified, as they account for all sources of information accessible to Bob, including the received message $A'$, his local share of the resource $B_l$, and any correlation between $X_0$ and $X_1$ that might be exploitable. Importantly, within this formulation, the obtained description avoids making any assumptions about the specific communication protocol employed, which weakens the operational criteria. As a result -- and to the best of our knowledge -- the criterion in Eq.~\eqref{eq:newic1} constitutes the most stringent information-theoretic principle currently available for bounding correlations in communication scenarios of the type represented in Fig.~\ref{fig:dag2}.

\begin{table}
  \centering
  \caption{Table containing the threshold values of $\gamma$ for the slices $\gamma\mathbf{p}_{\mathrm{Bell}}^{k} + (1-\gamma)\mathbf{1}$, where $\mathbf{p}_{\mathrm{Bell}}^{k}$ denotes the extremals from $(4,4;2,2)$ Bell scenario. In this case we present the values for quantum ($\gamma_Q$), IC ($\gamma_{\text{IC}}$) and novel informational criteria in Eq.\eqref{eq:newic1} ($\gamma_{\text{new}}$). The tuple $(j,l)$ specifies which form of \eqref{eq:newic1} yields the optimal bound on $\gamma$.}
  \label{table:noise_resilience}
  \setlength{\tabcolsep}{8pt}   
  \renewcommand{\arraystretch}{1.15} 

  \begin{tabular}{c |c c c c}
    \toprule
    Class. & ( $j,l$ ) & $\gamma_Q$ & $\gamma_{\text{IC}}$ & $\gamma_{\text{new}}$ \\
    \midrule
1 &  --  & 1.0 & 1.0 & 1.0 \\
2 & (1,1) & 0.707 & 0.943 & 0.894 \\
3 & (1,1) & 0.6 & 0.784 & 0.756 \\
4 & (1,0) & 0.6 & 0.894 & 0.816 \\
5 & (1,1) & 0.577 & 0.816 & 0.707 \\
6 & (1,1) & 0.707 & 0.816 & 0.816 \\
7 & (0,1) & 0.535 & 0.943 & 0.784 \\
8 & (1,1) & 0.541 & 0.853 & 0.853 \\
9 & (0,1) & 0.5 & 1.0   & 0.816 \\
10 & (0,0),(1,1) & 0.707 & 0.707 & 0.707 \\
11 & (1,1) & 0.707 & 0.943 & 0.894 \\
12 & (1,0) & 0.6 & 0.943 & 0.834 \\
13 & (0,0),(1,1) & 0.707 & 0.816 & 0.816 \\
14 & (0,1) & 0.577 & 1.0   & 0.816 \\
15 & (1,1) & 0.707 & 1.0   & 0.934 \\
16 & (1,1) & 0.707 & 1.0   & 0.894 \\
17 & (0,1) & 0.707 & 0.935   & 0.93  \\
18 & (0,1) & 0.6 & 1.0   & 0.93  \\
19 & (1,1) & 0.707 & 1.0   & 0.941 \\
20 & -- & 0.707 & 1.0 & 1.0 \\
    \bottomrule
  \end{tabular}
\end{table}

In Table~\ref{table:noise_resilience}, we show that the criterion in Eq.~\eqref{eq:newic1} outperforms all currently known operational formulations of IC in bounding correlations arising in the $(4,4;2,2)$ Bell scenario. Concretely, we introduce white noise into the extremal correlations of class $k\in\{1,2,\ldots,20\}$ in the form $ \gamma \mathbf{p}_{\mathrm{Bell}}^{k} + (1-\gamma)\mathbf{1}/4$ and report the minimum value of $\gamma$ for which a violation of the corresponding inequality is observed for some binary symmetric channel parametrized by $\epsilon\in[0,1/2]$. In contrast to the $(3,2;2,2)$ GC scenario, there exist classes for which the bounds implied by the entropic description coincide (up to numerical precision) with those allowed by quantum mechanics. This occurs for class 10. For the remaining extremal classes, although the quantum bound is not fully recovered, the inequalities derived from Eq.~\eqref{eq:newic1} significantly strengthen the limitations imposed by earlier criteria. Of particular importance, for classes 9, 14, 15, 16, 18, and 19, the previous operational formulations of IC did not reveal any implausible consequences, whereas Eq.~\eqref{eq:newic1} does identify. Despite this substantial improvement, there remain cases -- most notably classes 1 and 20 -- for which no violation is observed. Class 1 consists solely of classical extremals, and hence no violation is expected. Class 20, in turn, exhibits a pronounced gap between the quantum ($\gamma_Q$) and the NS bounds, yet neither the previously known criteria nor the newly established ones based on Eq.~\eqref{eq:newic1} succeed in identifying any implausibility for this class.

Importantly, given the generality of the introduced framework, the resulting bounds are entirely protocol-independent. Consequently, we have shown conclusively that, although the information-theoretic description of the causal structure in Fig.~\ref{fig:dag2} yields the most stringent form of \textit{communication-based} operational principle to date, it remains insufficient to fully reproduce the quantum nonlocality limits for the $(42;22)$ GC scenario. Indeed, by construction, our results are tightly connected to the pattern of nonlocal correlations achievable within the specific $(4,4;2,2)$ Bell scenario. It therefore remains possible that additional relevant information-theoretic constraints might emerge from the analysis of other Bell scenarios. Nevertheless, since the proposed method is explicitly optimized to bound correlations arising from the $(4,4;2,2)$ scenario, there is no compelling reason to expect that any such missing inequalities would strengthen the bounds presented in Table \ref{table:noise_resilience}.

\section{Discussion}

In this work, we introduced a new framework in which implausible consequences emerging from communication assisted by post-quantum resources can be systematically investigated for a broader range of communication tasks beyond those defined by the RAC and CC tasks. In this context, we propose that the soundness of the IC principle should extend to any scenario involving communicating parties, regardless of the specific task they aim to optimize. Within this framework, we have demonstrated how the current DI formalism of PM scenarios allows the identification of all relevant tasks in which nonlocal correlations may lead to such implausible consequences, such as violations of information-theoretic bounds of GC scenarios. Motivated by this fact, we have also presented a systematic search for the most suitable informational inequalities in the context of GC tasks, as permitted by the geometric informational-theoretic description of causal structures \cite{Chaves2, Yeung}.

Our findings show that even in the simplest GC scenario, where no straight correspondence with the standard RAC task exists, there are tasks in which implausibilities of supra-quantum correlations can still be identified within the information-theoretic framework. Notably, these effects are not captured by previous communication-based approaches, such as the (NTCC), highlighting the inequivalence of the identified tasks to both standard RAC and CC problems. Nevertheless, despite the generality of the causal model in Fig.~\ref{fig:dag}, the resulting information-theoretic description fail to fully recover the bounds implied by \eqref{eq:noisy_original} for the standard $2\mapsto1$ RAC task. This limitation arises because the causal structure in Fig.~\ref{fig:dag} does not assume any specific structure for the initial data set $X$, resulting in an informational description that bounds only the receiver’s potential knowledge about the two sources of uncertainty, namely $X$ and the noisy channel. This observation naturally motivates the explicit inclusion of additional sources of uncertainty in the causal description.

Indeed, the causal structure depicted in Fig.~\ref{fig:dag2}, in which the sender initially possesses two independent sets of data, $X_0$ and $X_1$, demonstrates a richer structure, leading to novel informational inequalities describing the primitive IC notion. Of particular relevance is the fact that we demonstrate that the previously known form of IC, given by inequality \eqref{eq:noisy_original}, constitutes a strictly weaker and particular instance of the information-theoretic description of Fig.~\ref{fig:dag2}, encoded in Eq.~\eqref{eq:newic1}. This establishes the internal consistency of our framework, as the latter fully recovers -- and significantly extends -- the existing state-of-the-art formulations of \textit{communication-based} principles. Furthermore, our results reveal the existence of more suitable information-theoretic constraints depending on the communication task considered. In the face of the problematic deep protocol-dependence of the previous methods attempting to explain quantum nonlocality, it is important to stress that the presented results carry no protocol-dependence since our analysis of correlations starts from the GC scenarios themselves.

Several promising avenues for future research naturally arise. One especially interesting further direction would be understanding whether including more independent sets of data may lead to a convergence of the bounds to the quantum one. Also, extending the analysis to GC tasks in multipartite scenarios is another natural further direction of investigation.

Acknowledgments:--  The authors thank Carlos Vieira for fruitful discussions and suggestions. This work is partially carried out under IRA Programme, project no. FENG.02.01-IP.05-0006/23, financed by the FENG program 2021-2027, Priority FENG.02, Measure FENG.02.01., with the support of the FNP. This work was partially supported by the Foundation for Polish Science (IRAP project, ICTQT, contract No. MAB/218/5, co-financed by EU within the Smart Growth Operational Programme). M.P. acknowledges support from the NCN Poland, ChistEra-2023/05/Y/ST2/00005 under the project Modern Device Independent Cryptography (MoDIC). This study was financed in part by the Coordenação de Aperfeiçoamento de Pessoal de Nível Superior - Brasil (CAPES) - Finance Code 001 - and by the Brazilian National Council for Scientific and Technological Development (CNPq) (INCT-IQ and Grant Number 316657/2023-9).
\bibliography{ref.bib}

\clearpage
\appendix
\onecolumngrid

\section*{Supplementary Material}

\section{Characterization (3,2,2)}\label{sec:characterization}

As introduced in the main text, when the parties are limited to classical resources, their probability distributions are described by the following classical causal model:
\ba{\label{eq:classical_app}
p(a,b|x,y) = \sum_{\lambda} p(\lambda) p(a|x, \lambda) p(b|a,y,\lambda).}
Since the response functions 
$p(a|x, \lambda)$ and $p(b|a,y,\lambda)$ can always be expressed as convex combinations of deterministic distributions, Eq. \eqref{eq:classical_app} admits the equivalent deterministic decomposition
\ba{\label{eq:classical_det}
p(a,b|x,y) = \sum_{\lambda} p(\lambda)\delta_{a,d_{A,\lambda},(x)}\delta_{b,d_{B,\lambda}(a,y)}}
where $d_{A,\lambda}(x)$, $d_{B,\lambda}(a,y)$ are deterministic functions of their respective inputs, fully specified by the hidden variable $\lambda$. It follows that the classical polytope described by Eq. \eqref{eq:classical_app} has $|\lambda|= |A|^{|X|} \cdot |B|^{|A|\cdot|Y|}$ extremal points. By enumerating all such vertices and applying facet-enumeration methods -- such as the algorithm implemented in PANDA \cite{panda} -- one can switch to the facet representation of the polytope, thereby identifying all inequalities that witness nonclassicality within the considered communication scenario.

For the simplest GC scenario with $|X| = 3$, $|Y| = 2$, $|B| = 2$, and $|A| = 2$, our analysis yields 576 Bell-type facets of the form:
\begin{align}\label{eq:Bell-type_app}
    \sum_{a,b,x,y} C_{ab}^{xy} \:p(a,b|x,y) \le \beta_C,
\end{align}
These fall into 14 equivalence classes under relabeling symmetries. Among them, 2 classes correspond to the trivial non-negativity constraints (\ie, $p(a,b|x,y)\ge 0$ for all $x\in\{0,1,2\}$ and $a,b,x,y\in\{0,1\}$). The remaining 12 classes have representatives of the following form:

\begin{align}\label{eq:13ineq}
p(01|00) -p(01|20) -p(10|21) -p(11|21) &\le 0
\quad (\text{class 1}) \nonumber\\
p(10|01) +p(11|11) &\le 1 \quad (\text{class 2}) \nonumber\\
p(01|00) -p(01|20) +p(10|01) +p(11|11) -p(11|21) &\le 1 \quad (\text{class 3}) \nonumber\\
p(10|01) -p(11|00) +p(11|10) -p(11|20) +p(11|21) &\le 1 \quad (\text{class 4}) \nonumber\\
p(01|00) +p(01|10) -p(01|20) +p(10|01) -p(10|21) +p(11|11) -p(11|21) &\le 1 \quad (\text{class 5}) \nonumber\\
p(01|00) -p(01|20) +p(10|01) -p(11|00) +p(11|01) +p(11|10) -p(11|20) &\le 1 \quad (\text{class 6}) \nonumber\\
p(01|00) +p(01|01) +p(01|10) -p(01|11) -p(01|20) +p(10|01) -p(10|21)\quad &\nonumber\\ +p(11|01) -p(11|21) &\le 1 \quad (\text{class 7}) \nonumber\\
p(01|00) +p(01|10) -p(01|20) +p(10|01) -p(10|21) -p(11|00) +p(11|01)\quad &\nonumber\\ +p(11|10) -p(11|21) &\le 1 \quad (\text{class 8})\nonumber\\
p(01|00) +p(01|01) +p(01|10) -p(01|11) -p(01|20) +p(01|21) +2p(10|01) \quad&\nonumber\\-p(10|11) -p(10|21)  +p(11|01) &\le 2 \quad (\text{class 9}) \nonumber\\
p(01|00) +p(01|10) -p(01|20) +2p(10|01) -p(10|21) -p(11|00) +p(11|01)\quad &\nonumber\\ +p(11|10)  +p(11|11)+p(11|20)  -2p(11|21) &\le 2 \quad (\text{class 10}) \nonumber\\
p(01|00) +p(01|01) -p(01|11) -p(01|20) +p(10|01) -p(10|21) +p(11|01) -p(11|11) &\le 1 \quad (\text{class 11})\nonumber\\
p(01|00) -p(01|20) +p(10|01) -p(10|21) -p(11|10) +p(11|11) +p(11|20) -p(11|21) &\le 1 \quad (\text{class 12})
\end{align}

\section{Details about the entropic characterization}\label{ap:entropic_characterization}

In this section, we provide a detailed information-theoretic analysis of each DAG introduced in the main text. We begin with the characterization of the purely classical case and subsequently show how these results extend to the quantum setting.

\subsection{Solving the GC DAG}

The scenario depicted in Fig.~\ref{fig:dag} of the main text involves the set of random variables $S=\{X, A, A', B_0, B_1, \mu_{\text{NS}}\}$, where $\mu_{\text{NS}}$ denotes the most general nonsignaling resource shared by the parties. This resource reduces to a quantum state in the quantum case ($\mu_{\text{NS}}\longrightarrow \rho_{\text{AB}}$) and to a classical variable when the parties are restricted to classical resources ($\mu_{\text{NS}}\longrightarrow \lambda$). As stated above, we first focus on the classical case and therefore assume $\mu_{\text{NS}}\longrightarrow \lambda$. Two structural features are central in this setting. First, since all variables in $S$ are jointly observable, the entire set $S$ forms a coexisting set. Second, the inequalities derived from weak monotonicity reduce here to the standard \textit{monotonicity} relation, $H(A,B)-H(B)\ge 0$ (vide Ref.~\cite{Chaves2}). In addition, classicality ensures that the causal relations are fully determined by the conditional independence relations (CIR), and therefore no data-processing inequalities are required. The only relevant CIRs in this case are $I(X:\lambda) =0$ and $I(A':X, \lambda| A) = 0$, with the corresponding marginal scenario of the form $S_{M} = \{\{X, A, A', B_y\}_y\}$. 

At this stage, we are ready to eliminate all entropy terms not contained in $S_{M}$ using the Fourier-Motzkin elimination method. The resulting information-theoretic description consists of 92 entropic inequalities. Among these, only 29 provide non-trivial constraints; the remaining ones follow directly from the basic Shannon-type inequalities and do not encode any restriction specific to the DAG under consideration. Remarkably, among these 29 inequalities, only the two given in the following inequalities (for $y \in \{0,1 \}$) are operationally relevant for bounding the correlations generated in the GC scenario, as summarized in Table~\ref{table:1} of the main text:
\begin{align}\label{eq:relevant_app}
I(X:A, A', B_y) + I(A: A', B_{y\oplus 1}) \le H(A) + I(A:A').
\end{align}

We now extend this description to the quantum case, which can be performed via linear-programming methods. The first step is to list all constraints associated with the respective coexisting sets, namely $\{X, A, A', B_0\}$, $\{X, A, A', B_1\}$, $\{X, A, A', \rho_{\text{B}}\}$ and $\{X, \rho_{\text{A}}, \rho_{\text{B}}\}$, where $\rho_{\text{A}}$ and $\rho_{\text{B}}$ denote the reduced states of the shared bipartite quantum state $\rho_{\text{AB}}$. In this setting, the only remaining causal information constraints (CIR) take the form:
\begin{align}
    I(X:\rho_{\text{A}}, \rho_{\text{B}}) = 0,\\
    I(A': X, \rho_{\text{B}}|A) = 0,\\
    I(X:B_y|A') = 0.
\end{align}
In this case, the causal relations must be supplemented by the data-processing relations, which have the form:
\begin{align}\label{eq:dataprocessing}
    I(X:A')&\le I(X:A), \\
    I(X:A',\rho_{\text{B}})&\le I(X:A,\rho_{\text{B}}), \\
    I(\rho_{\text{B}}:A)&\le I(\rho_{\text{B}}:X,\rho_{\text{A}}), \\
    I(\rho_{\text{B}}:X,A)&\le I(\rho_{\text{B}}:X,\rho_{\text{A}}), \\
    I(X:B_y)&\le I(X:A), \\
    I(X:B_y)&\le I(X:A,A'), \\
    I(X:A'B_y)&\le I(X:A), \\
    I(X:B_y)&\le I(X:A'), \\
    I(X:A',B_y)&\le I(X:A',\rho_{\text{B}}), \\
    I(X:A,A',B_y)&\le I(X:A,A',\rho_{\text{B}}), \\
    I(A:B_y)&\le I(A:A',\rho_{\text{B}}), \\
    I(X,A:A',B_y)&\le I(X,A:A',\rho_{\text{B}}), \\
    I(X,A:B_y)&\le I(X,A:A',\rho_{\text{B}}), \\
    I(X:A',B_y)&\le I(X:A,B_y). 
\end{align}

At this stage, all constraints can be arranged into a single matrix $M_Q$, which together with the $2^n$-dimensional entropy vector $\vec{H} = (H(\phi)),H(X_1), H(X_2), ..., H(X_1,X_2,...,X_{n-1}))$ defines the full set of linear information-theoretic constraints for the quantum case, \ie, $M_Q\cdot\vec{H}\ge \vec{0}$.  Moreover, any linear entropic inequality can be identified by the vector of coefficients $\vec{\mathcal{E}}$, so that the inequality takes the form of the inner product $\vec{\mathcal{E}}\cdot\vec{H}\ge 0$. Hence, the entropic inequality $\vec{\mathcal{E}}\cdot\vec{H}\ge 0$ is implied by the quantum information-theoretic description $M\cdot\vec{H}\ge \vec{0}$ if and only if the result of the linear program,
\begin{align}\label{lp:farkas}
    \underset{\vec{H}}{\text{minimize}} &\quad  \vec{\mathcal{E}}\cdot\vec{H}\\
    \text{subject to}&\quad M_Q\cdot\vec{H}\ge \vec{0},\nonumber
\end{align}
is precisely $\vec{\mathcal{E}}\cdot\vec{H}^* = 0$. This equivalence follows directly from Farkas' lemma (see Theorem 14.3 from Ref.~\cite{Yeung} and also Ref.~\cite{Chaves2}). Therefore, any candidate entropic inequality can be checked for validity within the quantum causal description $M_Q \cdot \vec{H} \ge \vec{0}$. Of particular relevance, this procedure confirms the validity of the inequality in Eq.~\eqref{eq:relevant_app} for the DAG depicted in Fig.~\ref{fig:dag}.

\subsection{`Three sources of uncertainty' DAG}

A similar procedure can be employed to obtain the information-theoretic description of the DAG depicted in Fig.\ref{fig:dag2} of the main text. In this case, the set of variables is $S=\{X_0, X_1, A, A', B_0, B_1, \lambda\}$, and the causal structure is fully characterized by the following CIRs:
\begin{align}
    I(X_0, X_1:\lambda) = 0,\\
    I(A': X_0, X_1, \lambda|A) = 0,\\
    I(B_y:X_0, X_1, A|A',\lambda) = 0,\\
    I(A',B_y: X_0, X_1|A,\lambda) = 0.
\end{align}
The corresponding marginal scenario is $S_{M} = \{\{X_0, X_1, A, A', B_y\}_y\}$, where the last missing step is performing the FM elimination. However, due to the doubly exponential growth of the FM method regarding the number of variables, this problem quickly becomes computationally challenging. Indeed, since each eliminated variable may generate a large number of new inequalities, beginning with the $2^7$-dimensional initial entropic cone already renders a direct FM elimination intractable. Nonetheless, this difficulty can be bypassed by introducing an additional refinement to the algorithm: the elimination step is applied only to the subset of inequalities that remain relevant for the final marginal scenario. This selective elimination dramatically reduces the computational overhead and enables the derivation of the desired information-theoretic description for the DAG in Fig.~\ref{fig:dag2} of the main text.

\subsubsection{Selecting relevant entropic inequalities}\label{ap:selection}

To identify the relevant inequalities describing the DAG of interest, we must recall that the purpose of the information-theoretic characterization is to derive operational principles capable of detecting implausible consequences arising from nonsignaling (NS) resources in communication scenarios. As discussed in the main text, the simplest Bell scenario associated with the DAG in Fig.~\ref{fig:dag2} of the main text is the $(4,4;2,2)$ scenario. Among the $20$ equivalence classes of extremal NS correlations, some lead to implausible behaviours that should be captured by the information-theoretic constraints. Consequently, such extremals can be employed to identify which are the relevant entropic inequalities that must be included in the characterization. 

In this context, we exploit the fact that certain NS extremals generate final GC statistics that would require more communication between the parties to be reproduced in the completely classical setting \footnote{See, for instance, the standard $2\mapsto 1$ RAC: a PR-box allows the receiver to perfectly recover both input bits, while any classical strategy would require sending $2$ bits.}. This observation can be equivalently reformulated in terms of the entropy vector. Let $\vec{H}(\mathbf{p}_{\text{Bell}}^k)$ denote the entropic vector produced by the extremal $\mathbf{p}_{\text{Bell}}^k$ through the probability distributions defined by:
\ba{\label{eq:behavior_ns_app}
p(a,b|x,y) = \sum_{y'} p_{\mathrm{Bell}}(a,b|x,(y, y'))\delta_{y'= a}.
}
For some extremals, one obtains $\vec{H}(\mathbf{p}_{\text{Bell}}^k)$ such that $I_k(A:A') = 1$, whereas the classical theory would require $I_C(A:A') > 1$. Thus, let $M_C \cdot \vec{H} \ge 0$ be the set of constraints implied by the basic inequalities and the CIRs for the classical description. We then consider the following linear program:
\begin{align}\label{lp:selection}
    \underset{\vec{H}}{\text{minimize}} &\quad  I(A:A') \nonumber\\
    \text{subject to}&\quad M_C^j\cdot\vec{H}\ge \vec{0}\\
        &\quad \vec{H}_{S'_M} = \vec{H}_{S'_M}(\mathbf{p}_{\text{Bell}}^k)\nonumber
\end{align}
where $M_C^j$ denotes the matrix $M_C$ with its $j$-th inequality removed, $\vec{H}_{S'_M} = \vec{H}_{S'_M}(\mathbf{p}_{\text{Bell}}^k)$ imposes equality constraints on the components of $\vec{H}$ belonging to the marginal scenario $S'_M = \{\{X_0, X_1, B_y\}_y\}$, and $\mathbf{p}_{\text{Bell}}^k$ is an NS extremal leading to the implausibility under analysis. By construction, replacing $M_C^j$ with the full matrix $M_C$ in \eqref{lp:selection} would necessarily yield $I_C(A:A') > 1$. Thus, if \eqref{lp:selection} returns a value $I^*(A:A')$ satisfying $I_C(A:A') > I^*(A:A') \ge 1$, the removal of inequality $j$ affects the classical bound and is therefore \emph{relevant} for the information-theoretic description. If, however, the result satisfies $I^*(A:A') = I_C(A:A')$, then inequality $j$ imposes no restriction on the classical performance and can be safely discarded.

Hence, the linear program \eqref{lp:selection} provides a method to identify which inequalities must be included in the characterization. The procedure is applied recursively: after testing inequality $j$, the matrix $M_C^j$ is updated -- either reinstating or permanently removing the $j$-th inequality -- and a new inequality is selected for the next round. After several iterations, the number of remaining inequalities is substantially reduced, and the Fourier-Motzkin elimination can then be efficiently applied to obtain the constraints over the marginal scenario $S_M$. The resulting inequalities necessarily detect the implausibility associated with the chosen extremal $\mathbf{p}_{\text{Bell}}^k$.

It is important to stress that this selection procedure is highly order-dependent. In particular, it may occur that a relevant inequality $l$ becomes irrelevant \emph{after} the removal of inequality $j$; in this case, $l$ will be excluded. For this reason, the algorithm must be executed multiple times -- with random ordering -- to ensure that the strongest informational constraint forbidding $\mathbf{p}_{\text{Bell}}^k$ is identified. In parallel to that, a stopping criterion is also required. For the fixed extremal $\mathbf{p}_{\text{Bell}}^k$, one may evaluate the robustness of the resulting inequalities by determining the threshold value $\alpha_E$ at which the inequality ceases to be violated over the mixture $\alpha \mathbf{p}_{\text{Bell}}^k + (1-\alpha)/4$. The algorithm may be halted whenever $\alpha_E$ fails to improve over several rounds. Naturally, this process must be performed for all classes of NS extremals to guarantee that all implausibilities of the given GC scenario are properly addressed.

Analogously to the previous case, the final step is to ensure that the resulting operational principles also hold in the quantum regime, using the linear programming method described in \eqref{lp:farkas}. The quantum description includes the coexisting sets: $\{X_0, X_1, A', A, B_0\}$, $\{X_0, X_1, A, A', B_1\}$, $\{X_0, X_1, A, A', \rho_{\text{B}}\}$ and $\{X_0, X_1, \rho_{\text{A}}, \rho_{\text{B}}\}$, with remaining CIR of the form
\begin{align}
    I(X_0,X_1:\rho_{\text{A}}, \rho_{\text{B}}) = 0,\\
    I(A':X_0,X_1, \rho_{\text{B}}|A) = 0,\\
    I(B_y: X_0,X_1|A') = 0.
\end{align}
The corresponding data-processing inequalities are derived from \eqref{eq:dataprocessing} by substituting $X$ with $X_0$, $X_1$, and $X_0X_1$ in all relations. With all constraints now encoded in the quantum description $M_Q \cdot \vec{H} \ge \vec{0}$, each derived inequality can be certified as valid for the quantum case.

\section{Method for obtaining entropic bounds}

In this section, we address how to compute the bounds implied by the new information-theoretic characterization for the communication tasks identified in Eq.~\eqref{eq:13ineq} in the simplest GC scenario. For convenience, we denote these inequalities by $E^{j}(\mathbf{p}_{\mathrm{Bell}})$, where the index $j$ labels the classes in \eqref{eq:13ineq}, and where each $E^{j}$ depends on the Bell correlations through \eqref{eq:behavior_ns_app}. As discussed in the main text, the NS bounds are determined by the nonsignaling vertices of the associated $(3,4;2,2)$ Bell scenario, consisting of 13952 vertices separated into 13 equivalence classes. Accordingly, the method relies on analyzing isotropic correlations of the form:
\ba{\label{eq:isotropic}\mathbf{p}_{\mathrm{Bell}}^{\alpha,k,l} \equiv \alpha \mathbf{p}_{\mathrm{Bell}}^{k,l} + (1-\alpha)\frac{1}{4},}
where $\alpha\in[0,1]$, and $\mathbf{p}_{\mathrm{Bell}}^{k,l}$ denotes the $l$-th extremal representative of class $k$. In this framework, the bound implied by the information-theoretic relations is obtained by solving
\begin{align}\label{opt:entropic_bound}
    \underset{\mathbf{p}_{\mathrm{Bell}}^{\alpha,k,l}}{\text{maximize}} &\quad  E^j(\mathbf{p}_{\mathrm{Bell}}^{\alpha,k,l}) \\
    \text{subject to}&\quad \mathcal{I}\cdot\vec{H}_{S_M}(\mathbf{p}_{\mathrm{Bell}}^{\alpha,k,l})\ge \vec{0},\nonumber
\end{align}
where matrix $\mathcal{I}$ contains all coefficients of the constraints arising from the final information-theoretic characterization. Thus, solving \eqref{opt:entropic_bound} amounts to finding $\alpha,k,l$ that maximize the task $E^{j}$ while respecting the information-theoretic principles encoded in $\mathcal{I}\cdot\vec{H}_{S_M}\ge \vec{0}$. Because NS vertices within the same class $k$ are equivalent under relabelings, any extremal $l$ can be transformed into another extremal $l'$ of the same class by local operations. This procedure should not produce a violation of the entropic inequalities, as prevented by data-processing. Therefore, for a fixed class $k$, the maximum allowed value of $E^{j}$ is given by the minimum over all $l$ of $E^{j}(\mathbf{p}_{\mathrm{Bell}}^{\alpha,k,l})$. Consequently, the global solution of \eqref{opt:entropic_bound} is of the form:  $\underset{k}{\max} \{ \underset{l}{\min} \{\underset{\alpha}{\max}  \{ E^j(\mathbf{p}_{\mathrm{Bell}}^{\alpha,k,l})\}\}\}$.

Notably, only inequality \eqref{eq:relevant_app} is violated in the simplest GC scenario through this procedure. Furthermore, in agreement with earlier results \cite{ICnoisy}, taking the limit $\epsilon \to 1/2$ yields the optimal values; accordingly, we fixed $\epsilon=0.50001$ in the calculations. The resulting bounds for each of the tasks in \eqref{eq:13ineq} are reported in Table~\ref{table:1} of the main text.

\end{document}